# Thickness Scaling Effect on Interfacial Barrier and Electrical Contact to Two-Dimensional MoS$_2$ Layers

Song-Lin Li,[1,2,†,*] Katsuyoshi Komatsu,[1] Shu Nakaharai,[1] Yen-Fu Lin,[3] Mahito Yamamoto,[1] Xiangfeng Duan,[4,5] and Kazuhito Tsukagoshi[1,*]

[1]*WPI Center for Materials Nanoarchitechtonics and* [2]*International Center for Young Scientist, National Institute for Materials Science, Tsukuba, Ibaraki 305-0044, Japan*

[3]*Department of Physics, National Chung-Hsing University, Taichung 40227, Taiwan*

[4]*Department of Chemistry and Biochemistry and* [5]*California Nanosystems Institute, University of California, Los Angeles, California 90095, USA*

\* Address correspondence to songlinli@gmail.com or tsukagoshi.kazuhito@nims.go.jp

**TOC graphic**

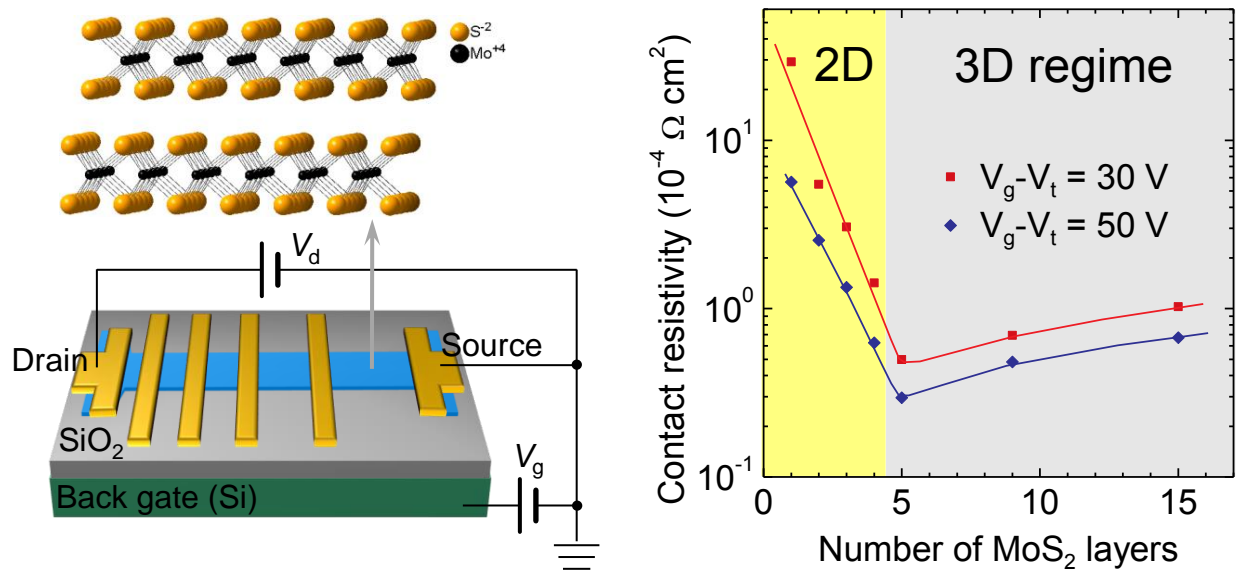

[†]Present address: ISIS & icFRC, Université de Strasbourg & CNRS, Strasbourg 67083, France

**ABSTRACT**

Understanding the interfacial electrical properties between metallic electrodes and low dimensional semiconductors is essential for both fundamental science and practical applications. Here we report the observation of thickness reduction induced crossover of electrical contact at Au/MoS$_2$ interfaces. For MoS$_2$ thicker than 5 layers, the contact resistivity slightly decreases with reducing MoS$_2$ thickness. By contrast, the contact resistivity sharply increases with reducing MoS$_2$ thickness below 5 layers, mainly governed by the quantum confinement effect. It is found that the interfacial potential barrier can be finely tailored from 0.3 to 0.6 eV by merely varying MoS$_2$ thickness. A full evolution diagram of energy level alignment is also drawn to elucidate the thickness scaling effect. The finding of tailoring interfacial properties with channel thickness represents a useful approach controlling the metal/semiconductor interfaces which may result in conceptually innovative functionalities.

Modern microelectronics roots in a fine control with gate bias on the height of potential barriers and the flow of charges at the interfaces between metallic contacts and active semiconductor channels,[1] which led to a great success of the semiconductor industry and revolutionized our life. The formation of ohmic contacts and high-efficient carrier transfer is the first step to construct high-performance devices.[2] Recently, layered transition-metal dichalcogenides (TMDs)[3-5] have attracted great interest not only for post-silicon electronics,[6-8] but also for optoelectronic[9-14] and photovoltaic[15,16] applications. The concurrence of atomic thickness and sizable bandgap promises them next-generation transistor channels after silicon. In addition, the exotic symmetry breaking in band structure, high optical absorption and mechanical flexibility can be exploited for valleytronic and photovoltaic devices. Undoubtedly, all the electrical systems begin with carrier transfer from electrodes to semiconductor channels; a profound understanding on the interfacial behavior between them is truly essential.[2]

In conventional bulk materials, the interfacial properties are basically independent on their dimensions. However, the physical scenario totally changes in the low dimensional systems. Generally speaking, the reduced material dimension increases bandgap ($E_g$) due to quantum confinement, a ubiquitous phenomenon in low-dimensional systems, such as quantum dots[17] and carbon nanotubes ($E_g \propto 1/\text{diameter}$).[18] The abnormal $E_g$ variation was extensively investigated in optical studies[19] but seldom studied in electrical experiments, although it is very important for contact design because an expanded $E_g$ may increase interfacial barrier height and suppress charge transfer at contacts, a potentially adverse factor for interfacial carrier injection. Such an assumption is confirmed in one-dimensional (1D) carbon nanotubes (CNTs) in which barrier height is reported to increase as tube diameter decreases.[18] In striking contrast, a recent study on two-dimensional (2D) $MoS_2$ seems to point to an opposite trend, showing that contact resistivity reduces in thinner layers.[20] To date, no consistent understanding has been reached regarding the dimension reduction effect on the contact properties in low-dimensional semiconductors. It deserves to point out that there are other efforts on the contact issue in 2D TMD semiconductors,[21-24] but the limited thickness range hinders a direct understanding the effect of thickness reduction on the material electrical properties. Actually, one of the advantages of TMDs lies in their atomic thickness which would allow the ultimate device downscaling in microelectronics. Hence, it is really important to know whether the reduced thickness is beneficial or detrimental to the contacts.

Herein, we perform a systematic thickness scaling study on the Au/2D $MoS_2$ interfaces based on a series of high-quality $MoS_2$ samples. The interfacial potential barrier is found to highly depend on $MoS_2$ thickness and increase from 0.3 to 0.6 eV as the $MoS_2$ thickness changes from 5 to 1 layer, as a result of quantum confinement. Similar to CNTs, reduced thickness of $MoS_2$ results in high interfacial barrier at Au/$MoS_2$ contacts. A linear correlation between barrier height and $MoS_2$ bandgap is revealed with a slope of *ca.* 0.5. A tentative full evolution diagram for energy level alignment is drawn to elucidate the thickness scaling effect on interfacial potential barrier. The thickness scaling rule adds fundamental knowledge in the interface physics and contact design for 2D semiconductors. In particular, the possibility in tailoring thickness interfacial potential barrier by channel thickness offers a useful way to interface engineering, which may find applications in functional devices such as tunneling transistors.[25,26]

**RESULTS AND DISCUSSION**

Although the contact issue of 2D TMDs has been addressed by several groups,[20-24] a systematic thickness scaling study remains absent. An underlying challenge for the thickness scaling study lies in the limited availability of large (>10 μm) TMD flakes because the conventional exfoliation approach based on direct TMD exfoliation between Scotch tapes and SiO$_2$ substrates[27] exhibits a considerably lower yield for TMD flakes (*ca.* 1 flake out of 10 times exfoliations) than that for graphene. In this regard, it is necessary to uncover the origin of the poor TMD exfoliation yields, in order to overcome the preparation barrier. We find that the poor TMD exfoliation yields originate from their unique surface condition. After initial thinning by Scotch tapes, the chalcogenide foils normally exhibit macroscopic ripples of tens to hundreds microns in length (Fig. S1a), in contrast to the flat surfaces exhibited by graphite foils (Fig. S1b). These ripples, which may arise from the reduced stiffness of TMDs, impede conformal adhesion of the foils to the target SiO$_2$ surfaces. Hence, adopting a viscoelastic medium in the last exfoliation step helps to increase the exfoliation yields. Here we adopt viscoelastic polydimethylsiloxane (PDMS) films as the supports to facilitate TMD exfoliation[28] (Fig. 1 a-d). Briefly, PDMS films are placed on glass slides using as exfoliation media, rather than rigid SiO$_2$ substrates, in the last exfoliation step. The few-layer TMD flakes are directly identified on the PDMS supports *via* optical contrast and are finally dry transferred to target SiO$_2$/Si substrates.

With optimizing exfoliation parameters, we managed to achieve large-area MoS$_2$ flakes in high yields, on average 1-2 flakes out of once exfoliation. Figure 1e shows an optical image of atomically thin 1- and 3-layer MoS$_2$ layers with a length of ~60 μm on a SiO$_2$/Si substrate. The thicknesses of the MoS$_2$ flakes are determined by the distance between the E$^1_{2g}$ and A$_{1g}$ Raman modes for thin flakes (Fig. 1f) as well as the intensity ratio between the MoS$_2$ and Si peaks for thick flakes.[29] To the best of our knowledge, they are among the largest TMD flakes prepared by mechanical exfoliation and are, so far, the largest TMD flakes used for studying metal/TMD contact issues.[20-24] The large sample size allows us to extract electrical parameters accurately, in contrast to the small samples used in previous reports. Also, the high exfoliation yields enable us to collect a series of flakes with consecutive numbers of layers, as shown in Fig. S2. The wide thickness distribution of the TMDs achieved enables us to investigate the thickness scaling effect of the metal/2D semiconductor.

The interesting few-layer $MoS_2$ flakes are normally connected to thick flakes. In effort to isolate them for electrical characterization, we identified efficient dry etchants ($CF_4$ and $CHF_3$) for patterning $MoS_2$ (Fig. 1h). The etching rate reaches 60 nm (~90 layers) per minute, which is more than 1000-fold higher than pure oxygen, the common etchant for graphene (Fig. 1i and S3). Additionally, $CF_4$ and $CHF_3$ are much cheaper than the early identified TMD etchant $XeF$,[30] representing an economic way for device fabrication.

Figure 2a illustrates pattern of the transfer line measurement. The $MoS_2$ flakes are top-contacted with multiple Au electrodes to extract line contact resistivity ($R_c$, in unit of Ω cm) and sheet resistance ($R_s$, in unit of Ω/square). After thermal annealing, a linear drain current ($I_d$) *versus* drain voltage ($V_d$) is observed (Fig. 2b), indicating an excellent contact between Au and $MoS_2$. No apparent current hysteresis is seen in the bi-directional $V_d$ scans, suggesting low trap states at the $MoS_2$/dielectric interfaces. The $MoS_2$ channels are meanwhile back-gated with a 285 nm $SiO_2$ to tune the channel carrier concentration. The capacitive coupling ability to $MoS_2$ channels is reflected in the transfer and output curves, which show an on/off current ratio of $10^7$ and clear current saturation at high $V_d$, respectively (top and bottom insets, Fig. 2b).

Figure 2c shows a typical transfer line plot for a bilayer (2L) sample under different gating conditions ($V_g$-$V_t$ from 10 to 50 V where $V_t$ is threshold voltage). The electrical parameters $R_c$ and $R_s$ are extracted from the intercepts and slopes of the linear fittings. The good linearity of the data points suggests high reliability of our data. Figure 2d summarizes $R_c$ *versus* gate voltage ($V_g$-$V_t$) for $MoS_2$ thickness from 5 to 1 layer. It is well known that the atomically thin flakes are extremely sensitive to gaseous absorbates and the $V_t$ position is mainly determined by annealing time and amount of absorbate remnants.[7,20] Despite an identical annealing time, large thickness-dependent $V_t$ positions are observed in our serial samples (Fig. S5), reflecting distinct electron trapping effects from the remnant surficial oxygen absorbates on samples with varied thickness. Hence, the inclusion of $V_t$ in the gating condition enables a fair comparison of the electrical behavior among devices.

Two features are shown in the $R_c$ curves. First, $R_c$ highly depends on gate voltage. When the gate bias increases from 0 to 50 V, $R_c$ is largely reduced by 10 to 50 folds depending on sample thickness. The large $R_c$ response to gate bias implies the presence of large tunneling current at the accumulation regime, a hint to elucidate the mechanism of charge injection. For our Au-contacted 5-layer thick sample, $R_c$ is 10 Ω cm at zero gate bias and is reduced to 1.0 Ω cm at 50 V gate bias, close to that reported in other groups.[21] Second, we confirm that the interfacial characteristics of 2D semiconductors follow the scaling rule of 1D CNTs in a similar way that reduced dimension leads to enhanced potential barrier,[18] since thiner $MoS_2$ flakes result in higher $R_c$ values. This observation implies that the previous report of lower $R_c$ in thinner samples[20] is likely a consequence of different extent of gold diffusion into underlying $MoS_2$ layers during long-time annealing, rather than the intrinsic electrical behavior of metal/semiconductor interfaces. In addition, as an advantage of the transfer line measurement, the intrinsic channel resistance $R_s$ can be also extracted together with $R_c$. Figure 2e shows $R_s$ *versus* gate voltage. Similar thickness dependence as $R_c$ is observed but they are of different origins. We have indicated previously that the strong thickness dependence of $R_s$ and carrier mobility (inset of Fig. 2e) is a natural result from the variation of Coulomb interaction distance between surficial charged impurities and channel carriers.[31]

With the presence of Schottky barrier at metal/1D CNT contacts, it is commonly accepted that CNT transistors operate as "Schottky barrier transistors" in which transistor action occurs primarily by varying the contact resistance rather than the channel conductance.[32] Such an operating mechanism has also been suggested in $MoS_2$ transistors.[23,24] To check how close the $MoS_2$ transistors are to Schottky barrier transistors, we calculate the $R_c$ percentage (ratio of $R_c$ to the total device resistance) for normalized $1\times1$ μm² square channels in Figure 2f. It is evident that the $R_c$ percentage, dependent on channel thickness, reaches 35-65% at the on transistor state. Besides, the $R_c$ percentage is larger in thinner device, indicating the increasingly influential role of contacts in the 2D materials. If considering a $10\times10$ nm² channel size for the post-silicon era, the $R_c$ percentage would approach 100% and completely dominate because $R_c$ increases 100 folds and $R_s$ is fixed. Then, the transistors evolve into a pure Schottky barrier transistor.

As far as electrical engineering is concerned, the area contact resistivity ($\rho_c$, in unit of Ω cm²) is commonly used to characterize contact quality, since it rules out the current crowding effect.[22] Here $\rho_c$ is derived from $R_c$ and $R_s$ by using the relation:[33]

$$R_c w = \sqrt{R_s \rho_c} \coth(d\sqrt{R_s/\rho_c}) \tag{1}$$

where $w$ is the channel width and $d$ is the contact length. Figure 3a shows the extracted $\rho_c$ values under different gate conditions for our samples, which exhibit similar gate dependence as $R_c$ and vary by more than 10 folds.

The availability of samples with consecutive numbers of layers enables a deep insight into the thickness scaling effect of 2D TMDs. A remarkable finding is emerged when we plot $\rho_c$ *versus* MoS$_2$ thickness (Fig. 3b). In contrast to the monotonic dependence of carrier mobility on thickness,[31] $\rho_c$ shows two opposite trends in different thickness regimes, with a positive slope in the 3D regime while a negative slope in the 2D regime, which forms a $\rho_c$ dip around 5 layers. As we will show, the formation of $\rho_c$ dip is a combined result of quantum confinement ($E_g$ modification) and the lopsided carrier distribution (thickness variation of inactive MoS$_2$ layer). Here the division of thickness regime, 2D or 3D regime, is justified by the $E_g$ magnitude in MoS$_2$. As MoS$_2$ thickness reduces from 5 to 1 layer (2D regime), $E_g$ expands from 1.2 to 1.8 eV[34,35] and $\rho_c$ increases by 10 folds under 50-V gating condition. The imitate relation between them suggests that the $E_g$ expansion changes the height of interfacial Schottky barrier accordingly. The presence of negative $\rho_c$ slope in the 2D regime is also consistent with that reported in 1D CNTs.[18] In the 3D regime ($\geq 5$ layers) where $E_g$ is fixed, $\rho_c$ is mainly determined by the thickness of the inactive middle MoS$_2$ layers[31,36] (Fig. 3c). The positive $\rho_c$ slope originates from the reduction of upper inactive layers as channel thickness reduces, which facilitates the carrier injection from electrodes to the lower active MoS$_2$ layers. On average, a reduction of one-layer thickness corresponds to a $\rho_c$ decrease of $\sim 4\times10^{-6}$ $\Omega$ cm$^2$. The positive $\rho_c$ slope behavior is nontrivial in the top-contacted transistors and has been studied in thick MoS$_2$ channels.[37] The transition of $\rho_c$ slope is a clear signature of dimensionality crossover, which can be used as a direct dimensionality criterion for distinguishing low-dimensional and bulk materials.

As Schottky barrier transistors,[23] the transistor switching states are determined by the barrier width, which is mainly modulated by gate bias and carrier concentration ($n$), through tuning the positions of channel energy levels, as shown in Fig. 3d. At low gate bias, charge injection is controlled by a thermal emission (TE) process nearly without tunneling component, which gives rise to the off transistor state. Since a slight increase of carrier concentration in the off transistor state does not obviously modify the barrier width or introduce tunneling current, theoretically the TE process leads to gate bias (equivalently carrier concentration) independent $\rho_c$ behavior following the relation[1,38]

$$\rho_c = \frac{k}{A^*Tq}\exp\left(\frac{q\phi_B}{kT}\right) \quad (2)$$

where $k$, $T$, $q$ and $\phi_B$ are the Boltzmann constant, temperature, elementary charge and interfacial Schottky barrier, the Richard constant $A^* = 8\pi m^* q k^2 h^{-3}$ with $m^*$ the effective mass and $h$ the Planck constant. In contrast, at high gate bias, the induced dense carriers considerably reduce the barrier width and increase the tunneling probability. Then, thermally assisted tunneling (also called thermal field emission, TFE) current populates channels, leading to the on transistor state. In this case, carrier injection shifts to the TFE mechanism and $\rho_c$ follows[1,38]

$$\rho_c = \frac{k\sqrt{E_{00}}\cosh(E_{00}/kT)\coth(E_{00}/kT)}{A^*Tq\sqrt{\pi q(\phi_B - u_f)}} \exp\left(\frac{q(\phi_B - u_f)}{E_{00}\coth(E_{00}/kT)} + \frac{qu_f}{kT}\right) \quad (3)$$

where $u_f$ is chemical potential and $E_{00} = qh\sqrt{n_{3D}/(4m^*\varepsilon)}$ is a doping related parameter with $\varepsilon$ the permittivity. Then, $\rho_c$ becomes highly dependent on carrier concentration (*i.e.*, gate bias) in this regime, which offers us a convenient way to estimate the values of barrier height $\phi_B$.

In 2D regime, the channel thickness is smaller than the screening depth[37] and the inactive layer is negligible. By assuming $n_{3D} = n_{2D}/\text{thickness}$, we find that all our $\rho_c$ data are well fitted to the TFE theory using Equation 3. As evident in Fig. 3e, a reasonable agreement between the experiment and calculation is reached. Note that large deviation at low gate regime ($V_g - V_t < 20$ V) appears in the monolayer and bilayer samples, which is attributed to the strong nonlinear dependence of *n* on gate bias around $V_t$, as shown in Fig. S5. In the low gating regime, the doping levels are noticeably underestimated by simply using a linear *n*-$V_g$ relation due to strong surficial doping effect from the residual absorbates. This strong doping effect arises from the enhanced surface/volume ratios and is commonly observed in other groups.[22,23] For this reason, the $\rho_c$ values at higher gate bias reflect $\phi_B$ more intrinsically. The biasing condition of $V_g - V_t = 50$ V is thus adopted to extract $\phi_B$.

Figure 3f plots the derived $\phi_B$ *versus* channel $E_g$ for all samples ranging from 1 to 5 layers. A linear fit reveals a $d\phi_B/dE_g$ slope of 0.46, which indicates that nearly half of the $E_g$ expansion due to thickness reduction is used to build up the interface barrier $\phi_B$. Note that the slope approaches 0.5, possibly suggesting that the $E_g$ expansion is rather symmetric relative to the Fermi level of Au electrode. In other words, the upshift of conduction band $E_c$ is approximate to the downshift of $E_v$. Such a behavior resembles that observed in 1D CNTs,[18] suggesting similar interfacial equilibrium dynamics between 2D and 1D semiconductors.

Figure 3g depicts a tentative evolution diagram for energy level alignment to summarize the thickness scaling effect on interfacial potential barrier, which represents fundamental knowledge for contact design for electrical devices based on 2D semiconductors. In modern microelectronics, $\rho_c$ is required to be as low as $10^{-8}\ \Omega\,cm^2$ to endure device miniaturization in deep sub-micrometer technology nodes. However, $\rho_c$ of the Au/MoS$_2$ contact is as high as $1\times10^{-4}\ \Omega\,cm^2$, about 4 orders of magnitude higher than the upper limit in microelectronics. Developing effective strategies to create more transparent contacts is necessary to employ the atomically thin semiconductors as post-silicon transistor channels.[6] Above systematic understanding on electrical contact to MoS$_2$ flakes provide useful guidance for contact design for devices based on 2D semiconductors. First, formation of narrow band gap or metallic sulfides such as TiS$_2$ at contact interfaces would be beneficial in reducing interfacial barrier, considering 4 orders lower $R_c$ shown by Pd/metallic graphene contacts.[39] Additionally, contact engineering with degenerate doping would further lower barrier width and result in efficient charge injection. Second, semiconductors with low bulk $E_g$ should be considered as channel candidates, given the fact of $E_g$ expansion after thinning down. In this sense, the search of different channel materials with technologically suitable $E_g$ is necessary.

On the other hand, the finding of tailoring interfacial properties with channel thickness also represents a useful approach that can control the metal/semiconductor interfaces, which may result in conceptually innovative functionalities. For instance, the barrier height is one of the most important parameters in the design of tunneling transistors.[25,26] A suitable barrier value has to be careful chosen to compromise the leakage and operating currents. Thus, the finding of finely controlling the barrier height with channel thickness offers a facial way to realize viable tunneling transistors.

**CONCLUSIONS**

We performed a systematic thickness scaling study on electrical contact to 2D semiconductors. A generalized interfacial dynamics is revealed in the 2D and 1D low-dimensional structures. As semiconductor thickness reduces, the Fermi level of contacting metals is strongly pinned due to the presence of large interface states. A large tenability of interfacial barrier spanning from 0.3 to 0.6 eV is observed when merely varying $MoS_2$ thickness from 5 to 1 layer. Thermal field emission is revealed as the responsible carrier injection mechanism, where carrier transfer relies on thermally assisted tunneling at metal/2D semiconductor interfaces. A detailed energy level alignment diagram is also established for different $MoS_2$ thicknesses. Our in-depth results offer insight into the interfacial electrical properties of 2D semiconductors, which would be beneficial for device design and performance optimization in electronic devices based on 2D semiconductors.

EXPERIMENTAL SECTION

The large $MoS_2$ flakes used in experiment were prepared from natural molybdenite crystals (Furuchi, Japan) by an improved mechanical exfoliation approach using viscoelastic PDMS as supports. The adoption of the viscoelastic supports and the formation of good conformal adhesion of TMD flakes to PDMS are critical for obtaining large flakes and high yields. Dry etching is performed under a fluorinated plasma environment with supplying a mixture of $CF_4$ (or $CHF_3$) and $O_2$ gases as etchants. A low ratio of $O_2$ was employed to prevent the generation of solid fluorocarbon residues, which were often observed after removing the resist masks and collapsed onto the target TMD flakes if no $O_2$ was introduced. All samples were annealed in flowing $H_2$/Ar gas at 300 °C for 2 hours before electrical characterization. All the electrical characterizations were performed at room temperature and vacuum surrounding (~$3 \times 10^{-4}$ Pa) in a probe station. An Agilent 4156C semiconductor analyzer was used for electrical characterization.

*Supporting Information Available:* Experimental details on sample preparation, dry etching, device fabrication, and theoretical analysis of current distribution at probe/channel contacts. This material is available free of charge *via* the Internet at http://pubs.acs.org.

*Acknowledgement.* This research was supported by a Grant-in-Aid (Kakenhi No. 25107004) from the Japan Society for the Promotion of Science (JSPS) through the Funding Program for World-Leading Innovative R&D on Science and Technology (FIRST), and Experiment-Theory Fusion trial project by WPI-MANA.

Figure and legend

# Figure 1

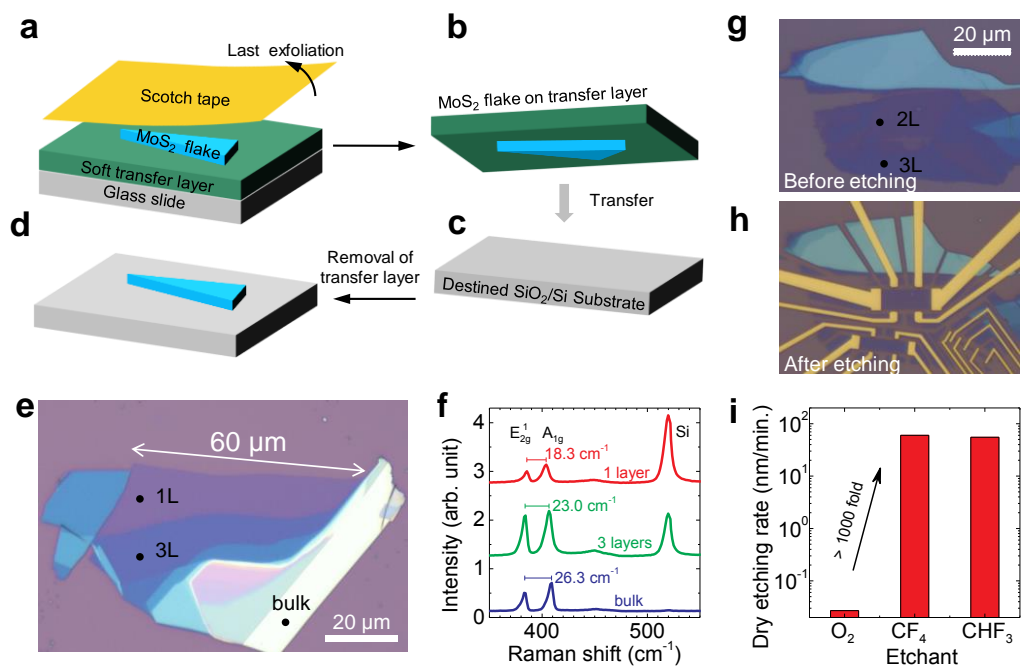

**Figure 1.** (a–d) Preparation flow of the mechanical exfoliation based on viscoelastic PDMS supports. (e) Optical images for as-transferred large $MoS_2$ flakes. (f) Typical Raman spectra for the $MoS_2$ flakes with different thicknesses. The flake thickness is reflected in the distances between $MoS_2$ $E_{2g}^1$ (~383 cm$^{-1}$) and $A_{1g}$ (~408 cm$^{-1}$) modes and the intensity ratio of the Si 520 cm$^{-1}$ mode to $MoS_2$ modes. (g) and (h) Samples before and after dry etching. (i) Comparison of etching rate for different etchants for $MoS_2$.

# Figure 2

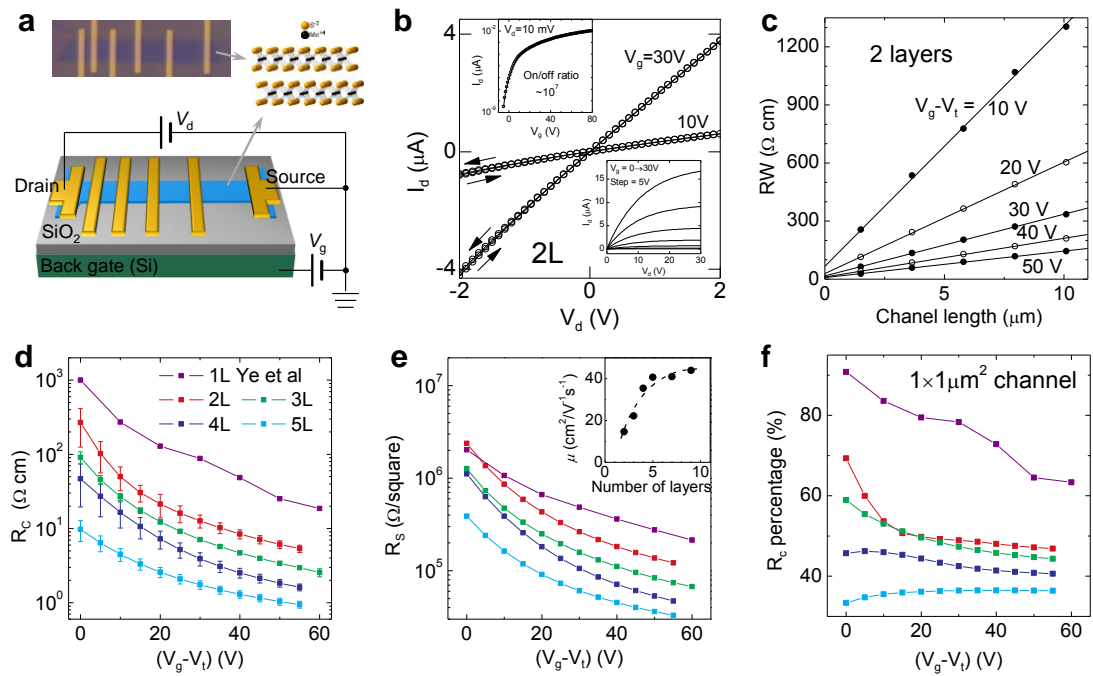

**Figure 2.** (a) Schematic diagram and real optical image for the geometry of transfer line measurement. The inset shows the atomic structure of $MoS_2$. (b) Typical electrical properties for bilayer $MoS_2$ field-effect transistors. Top and bottom insets are the corresponding transfer and output curves. (c) Transfer line plot for extracting line contact resistivity ($R_c$) and sheet square resistivity ($R_s$) under different gating conditions. (d) and (e) Extracted $R_c$ and $R_s$ for different sample thicknesses. Inset in (e): Carrier mobility ($\mu$) *versus* channel thickness. (f) Calculated ratios of $R_c$ to total device resistance for $1\times 1$ $\mu m^2$ square channels.

# Figure 3

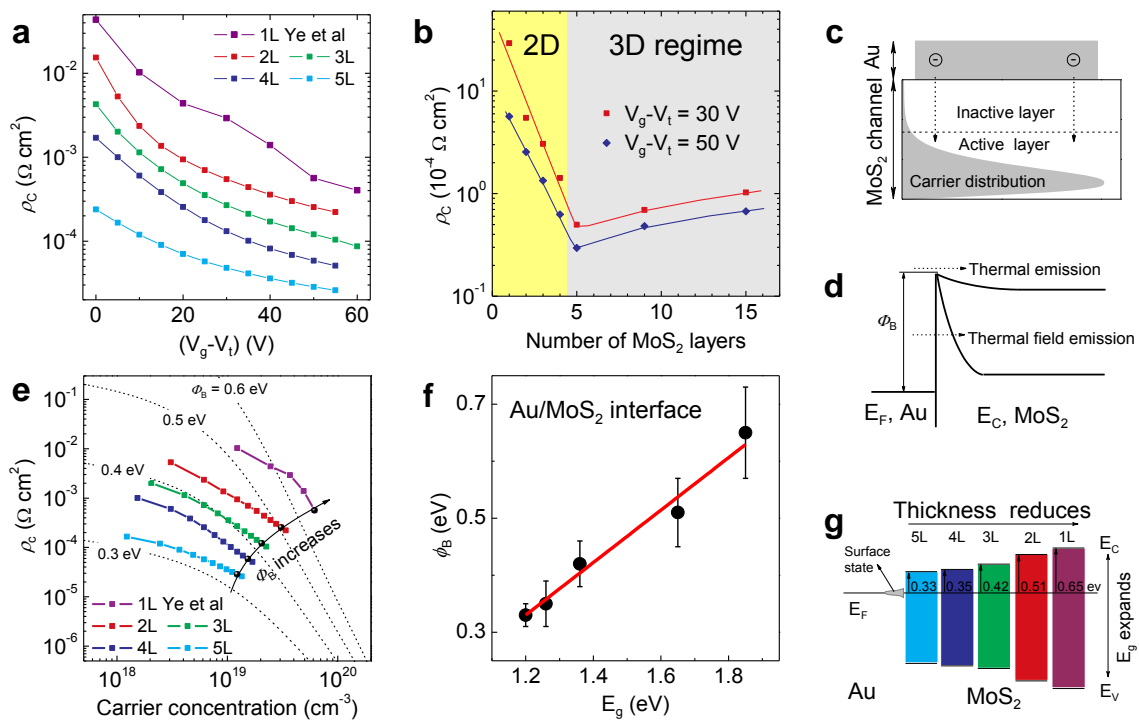

**Figure 3.** (a) and (b) Gate and thickness dependence of area contact resistivity ($\rho_c$). (c) Schematic carrier distribution and injection path for back-gated thick devices in the 3D regime. (d) Schematic diagram of band alignments for two carrier injection theories: thermal emission (TE) and thermal field emission (TFE). The difference between them lies in the width of interfacial barrier which changes with the gate bias and the carrier density in semiconductors. (e) Comparison of $\rho_c$ data (dotted lines) with theoretical results of TFE conduction mechanism (dashed lines) to extract barrier heights. (f) Thickness scaling effect on the interfacial barrier height ($\phi_B$) at Au/MoS$_2$ contacts, which is a function of semiconductor bandgap ($E_g$). (g) Evolution of energy level alignment at Au/MoS$_2$ interfaces as MoS$_2$ thickness reduces.

20